\documentclass[runningheads]{llncs}
\usepackage[T1]{fontenc}
\usepackage{graphicx}
\usepackage{xspace}
\usepackage{url}
\usepackage{color}

\newcommand{\rev}{\ensuremath{\mathrm{rev}}}
\newcommand{\kebab}{\texttt{kebab}\xspace}
\newcommand{\ropebwt}{\texttt{ropebwt3}\xspace}
\newcommand{\tagger}{\texttt{tagger}\xspace}
\newcommand{\bmove}{\texttt{b-move}\xspace}

\begin{document}

\title{KeBaB: $k$-mer based breaking\\
for finding long MEMs}
\titlerunning{KeBaB: $k$-mer based breaking}

\author{Nathaniel K.\ Brown\inst{1}\orcidID{0000-0002-6201-2301} \and \\
Lore Depuydt\inst{2}\orcidID{0000-0001-8517-0479} \and \\
Mohsen Zakeri\inst{1}\orcidID{0000-0002-9856-719X} \and \\ 
Anas Alhadi\inst{3} \and Nour Allam\inst{3} \and Dove Begleiter\inst{3} \and \\
Nithin Bharathi Kabilan Karpagavalli\inst{3} \and \\
Suchith Sridhar Khajjayam\inst{3} \and Hamza Wahed\inst{3} \and \\
Travis Gagie\inst{3}\orcidID{0000-0002-1825-0097} \and \\
Ben Langmead\inst{1}\orcidID{0000-0003-2437-1976}}

\authorrunning{N.K.\ Brown et al.}
\institute{Johns Hopkins University, Baltimore, USA \and
Ghent University, Ghent, Belgium \and
Dalhousie University, Halifax, Canada}

\maketitle
\setcounter{footnote}{0}

\begin{abstract}
Long maximal exact matches (MEMs) are used in many genomics applications such as read classification and sequence alignment. Li's ropebwt3 finds long MEMs quickly because it can often ignore much of its input.  In this paper we show that a fast and space efficient $k$-mer filtration step using a Bloom filter speeds up MEM-finders such as ropebwt3 even further by letting them ignore even more.  We also show experimentally that our approach can accelerate metagenomic classification without significantly hurting accuracy.

\end{abstract}

\section{Introduction}
\label{sec:intro}

A challenge for today's string-matching algorithms is to compute exact matches with respect to an index over a large, repetitive text.  This is a pressing problem in computational genomics, where databases of reference genomes and pangenomes are growing very rapidly.  
One highly practical full-text indexing method for pangenomes is \ropebwt\cite{Li24}, which indexes using a run-length compressed form of the Burrows-Wheeler Transform of the text.  Its strategy for querying the index involves skipping along the query in the style of Boyer-Moore pattern matching~\cite{BM77}, an idea that was first connected to BWT queries by Gagie~\cite{Gag24}.
Internally, \ropebwt uses a bidirectional FM index together with a forward-backward matching algorithm for finding long maximal exact matches (MEMs).

\begin{figure}[t]
    \centering
    \includegraphics[width=.9\linewidth]{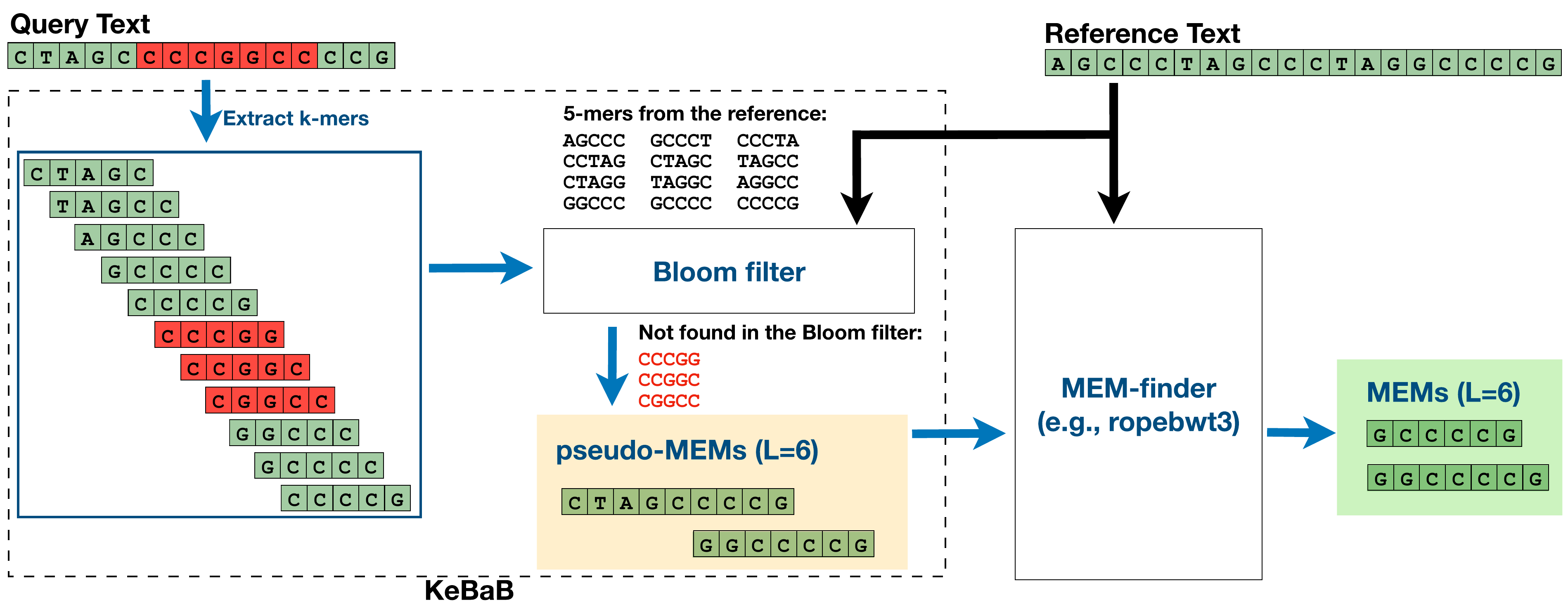}
    \caption{An example of how to use KeBaB to find pseudo-MEMs.}
    \label{fig:example}
\end{figure}

In this paper we propose a fast $k$-mer filtration strategy using a Bloom filter that allows for more skipping and speeds \ropebwt up substantially.  We call our strategy KeBaB for ``$k$-mer based breaking''.
In Section~\ref{sec:BML} we briefly review MEM-finding.  In Section~\ref{sec:breaking}, we describe how to break a pattern into substrings we call pseudo-MEMs that are guaranteed to contain all sufficiently long MEMs of the pattern with respect to an indexed text.  If we are interested only in the $t$ longest MEMs, then we can search in the pseudo-MEMs in non-increasing order by length and stop when we have found $t$ MEMs at least as long as the next pseudo-MEM.  This should require modifying \ropebwt but our experiments in Section~\ref{sec:experiments} indicate that simply searching in the $t$ longest pseudo-MEMs and discarding the rest does not significantly affect downstream results --- even compared to using all the long MEMs.  Figure~\ref{fig:example} shows an example of how to use KeBaB to find pseudo-MEMs.

\section{MEMs, forward-backward and BML}
\label{sec:BML}

A {\em maximal exact match} (MEM) --- also called super-maximal exact matches (SMEMs) --- of a pattern $P [0..m - 1]$ with respect to a text $T [0..n - 1]$ is a substring $P [i..j]$ such that
\begin{itemize}
\item $P [i..j]$ occurs in $T$,
\item $i = 0$ or $P [i - 1..j]$ does not occur in $T$,
\item $j = m - 1$ or $P [i..j + 1]$ does not occur in $T$.
\end{itemize}
Finding MEMs is an important step in many bioinformatics pipelines, such as aligning long and error-prone DNA reads to large pangenomic references.

For Li's~\cite{Li12} popular forward-backward MEM-finding algorithm, we keep FM-indexes~\cite{FM05} of $T$ and its reverse $T^\rev$.
Assuming all the characters in $P$ occur in $T$, the leftmost MEM starts at $P [0]$.  We can therefore find the leftmost MEM $P [0..e_1]$ by searching for $P^\rev$ in the index for $T^\rev$.
If $e_1 < m - 1$ then the second MEM $P [s_2..e_2]$ from the left in $P$ includes $P [e_1 + 1]$.  By definition, no MEM includes $P [s_2 - 1..e_1 + 1]$, so we can find $s_2$ by searching for $P [0..e_1 + 1]$ in the index for $T$.  Conceptually, we can then recurse on $P [s_2..m - 1]$ and find $e_2$ and the remaining MEMs.  The number of backward steps this takes in the indexes is proportional to the total length of the MEMs.

For many applications we are interested only in long MEMs, which are biologically significant since they are unlikely to be the result of noise.  Unfortunately, the total length of the MEMs is often dominated by many short MEMs, which we would like to ignore.  Suppose we are interested only in MEMs of length at least $L$.  Gagie~\cite{Gag24} recently observed that any such MEM starting in $P [0..L - 1]$ includes $P [L - 1]$, so if we search for $P [0..L - 1]$ in the index for $T$ and find that $P [s..L - 1]$ occurs in $T$ but $P [s - 1..L - 1]$ does not, for some $s > 1$, then we can ignore $P [0..s - 1]$ and recurse on $P [s..m - 1]$.  If we find that all of $P [0..L - 1]$ occurs in $T$ then we can still use the first few steps of forward-backward to find the leftmost MEM and the starting position of the second MEM from the left in $P$, and then recurse.  Since this approach is reminiscent of Boyer-Moore pattern matching, we call it {\em Boyer-Moore-Li} (BML).  Li~\cite{Li24} incorporated BML into \ropebwt and found it significantly accelerates MEM-finding.

\section{$k$-mer based breaking into pseudo-MEMs}
\label{sec:breaking}

Another technique for speeding up pattern matching is {\em $k$-mer filtration}.  In contrast to BML, this requires
scanning the whole input and deciding which parts can be ignored because they cannot contain significant-length matches.  If the alphabet's size is polylogarithmic in $n$ and BML uses a sublinear number of backward steps, then in the word-RAM model filtration is asymptotically slower;
however, the filtration scan is sequential, incurring few cache misses and allowing it to be fast in practice compared to FM-index queries, which tend to incur many cache misses.

Suppose we are given $k$ when we index $T$ and we build a Bloom filter~\cite{Blo70} for the distinct $k$-mers in $T$.  Bloom filters can give false-positive results but not false-negative ones, so if the filter answers ``no'' for a $k$-mer $P [i..i + k - 1]$ then no MEM of length at least $k$ includes that $k$-mer.  It follows that when we are given $P$ and $L > k$, we can break $P$ up into maximal substrings --- which can overlap by $k - 2$ characters but cannot nest --- containing only $k$-mers for which the filter answers ``yes'', that contain all the MEMs of length at least $L$.  We call these substrings \textit{pseudo-MEMs} because they are our best guesses at the MEMs of length at least $L$ based on the information we can glean from the filter.

\begin{definition}
A {\em pseudo-MEM} of a pattern $P [0..m - 1]$ with respect to a text $T [0..n - 1]$, an integer $k \geq 1$, a given Bloom filter for the distinct $k$-mers in $T$ and an integer $L > k$, is any maximal non-empty substring $P [i..j]$ of $P$ of length at least $L$ such that all the $k$-mers in $P [i..j]$ appear in the filter.
\label{def:pseudo_MEM}
\end{definition}

\begin{proposition}
All the MEMs of $P$ with respect to $T$ of length at least $L > k$ are contained in the pseudo-MEMs of $P$ with respect to $T$ and any Bloom filter for the distinct $k$-mers in $T$.
\end{proposition}

Our experiments in Section~\ref{sec:experiments} show that computing the pseudo-MEMs and searching in them is in practice already faster than searching in all of $P$.  Further, they show that if $T$ is highly repetitive then the Bloom filter tends to be smaller than the FM-indexes for \ropebwt.

If we seek only the top-$t$ longest MEMs of length at least $L$, however, then we can search the pseudo-MEMs in non-increasing order by length and stop when we have found $t$ MEMs at least as long as the next pseudo-MEM.  We can compute and sort the pseudo-MEMs independently of the actual MEM-finding algorithm we are using, but having it keep track of $t$ longest MEMs it has found and stop when the next pseudo-MEMs is shorter should require us to modify it. We have not yet done this for \ropebwt.

\begin{proposition}
If we seek only the top-$t$ longest MEMs of length at least $L$ and we are searching the pseudo-MEMs in non-increasing order by length, we can stop when we have already found $t$ MEMs longer than the next pseudo-MEM.
\end{proposition}

Without modifying \ropebwt, we can estimate how long it would take to find the top-$t$ MEMs by finding them ourselves ahead of time and giving \ropebwt only the pseudo-MEMs it would search in before stopping.  Our experiments in Section~\ref{sec:experiments} show that for reasonable values of $t$, this should be much faster than running \ropebwt on all the pseudo-MEMs; moreover, at least for the metagenomic classifier we tested, it does not significantly hurt the accuracy.  In fact, we found that using the long MEMs we found in only the top-$t$ pseudo-MEMs --- which are not guaranteed to be the top-$t$ MEMs but which we can find without modifying MEM-finders such as \ropebwt~--- is even faster and still results in nearly identical classification accuracy.

\section{Experiments}
\label{sec:experiments}

Our \kebab implementation in \texttt{C++} is available at \url{github.com/drnatebrown/kebab}. It streams over $k$-mers using a rolling nucleotide hash defined by ntHash supporting both forward and reverse complement~\cite{MCVB16}.  We use HyperLogLog~\cite{HLL07} to estimate the cardinality of a text collection to initialize the Bloom filter size, which is optimized with respect to the number of filter hashes used. We then add canonical $k$-mers (the smaller of each $k$-mer and its reverse complement by hash value) to the filter. Given a pattern, we query its canonical $k$-mers and extract the pseudo-MEMs.  We leave optimization details to the appendix.

\subsection{MEM-finding}

We tested the speed of MEM-finding on a mock community dataset of 7 microbial species ($5867$ genomes, $\sim\!27$ GB) from Ahmed et al.'s~\cite{ARGBL23} SPUMONI 2 study. Patterns consist of long ONT \textit{null reads} (10245 yeast reads with average length $19693$) and \textit{positive reads} (581802 microbial reads with average length $25378$). Constructing \ropebwt took $162.88$ minutes with an $0.7988$ GB index. Building \kebab with $k = 20$ and one hash function took $4.02$ minutes with an $0.2684$ GB filter (about a third of the size of \ropebwt's index).

\begin{figure}[t]
    \centering
    \includegraphics[width=.9\linewidth]{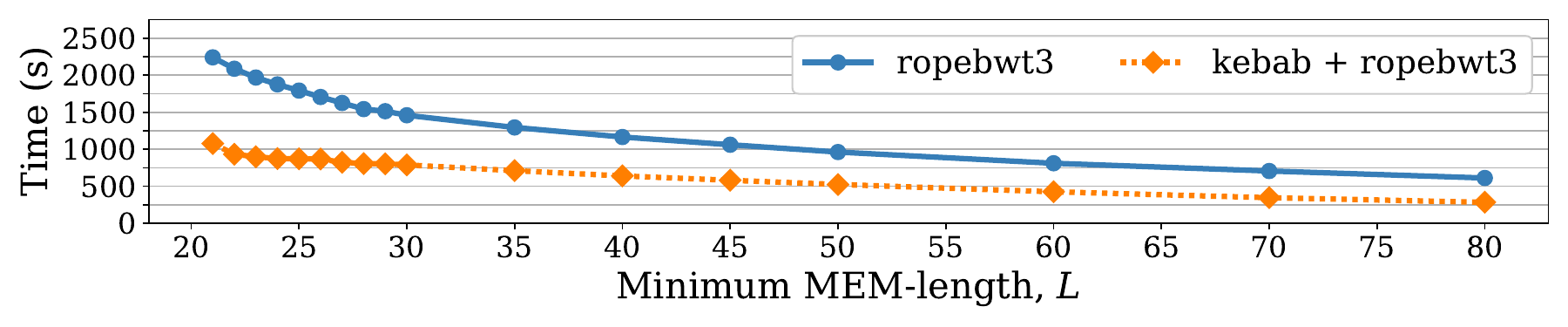}
    \caption{Total runtime in seconds for MEM-finding methods, searching in a microbial pangenome with different minimum MEM-length values $L$.}
    \label{fig:mem_length_experiment}
\end{figure}

\begin{figure}[t]
    \centering
    \includegraphics[width=.9\linewidth]{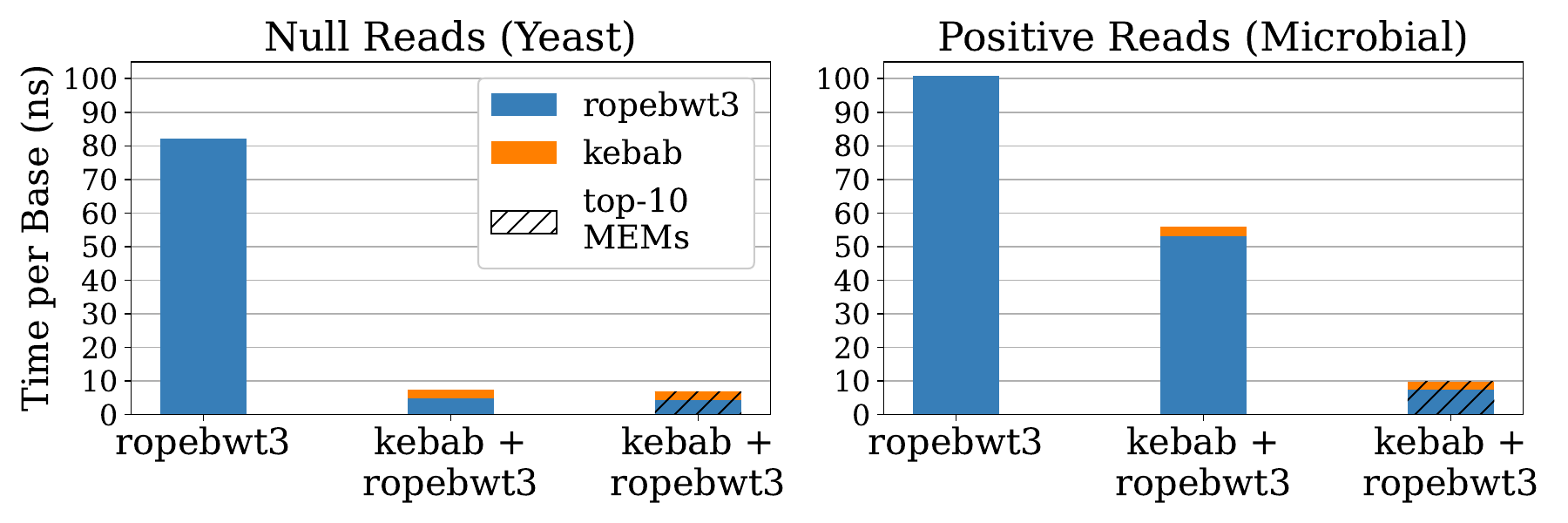}
    \caption{For $L=40$, time per base to find all long MEMs or only the 10 longest MEMs.}
    \label{fig:binary_experiment}
\end{figure}

We compared the time to find MEMs with \ropebwt alone with default settings, to the time to first generate pseudo-MEMs with \kebab and then search them with \ropebwt. We also simulated early stopping to find the 10 longest MEMs as explained in Section~\ref{sec:breaking}.
Figure~\ref{fig:mem_length_experiment} shows the total times for different choices of $L$, and Figure~\ref{fig:binary_experiment} shows times for null and positive reads with $L=40$. For $L\geq30$, the running-time of only the \kebab step on the reads was at most about 3 times more than the time to copy them to another file, which is a rough lower bound on file I/O for a filtering step.  

\subsection{Metagenomic Classification}

To see how using only a few long MEMs affects downstream applications, we replicated the metagenomic classification experiment in Depuydt et al.'s~\cite{DAFLG25} \tagger study, consisting of 8 microbial species ($8165$ genomes, $\sim\!37$GB) and $50000$ simulated long ONT reads with average length $5236$. By default, \tagger uses Depuydt et al.'s~\cite{DRVVGF25} bidirectional r-index \bmove with BML to find long MEMs together with sample species containing occurrences of them, then classifies the reads based on the sample species containing each read's long MEMs.  We note that \bmove is usually larger but faster than \ropebwt, so speedups with \kebab are not as dramatic.

We computed \tagger's accuracies (on the left) --- that is, its percentages of true-positive classifications --- and the average number of steps \bmove takes (on the right) when finding and classifying based on
\begin{itemize}
\item all the reads' MEMs of length at least $L$ (``default''),
\item only the longest $t$ MEMs from each read (``top-$t$ MEMs''),
\item only the MEMs of length at least $L$ in the longest $t$ pseudo-MEMs from each read (``top-$t$ pseudo-MEMs''),
\end{itemize}
for $L = 25$ and various values of $t$.  We ran \tagger with default settings and $L = 25$ because Depuydt et al.\ found it gave good results.  Clearly, for $t$ greater than about 10, using only the $t$ longest MEMs in each read or the MEMs of length at least $L = 25$ in the $t$ longest pseudo-MEMs, does not noticeably hurt \tagger's accuracy but significantly reduces the number of steps \bmove takes for MEM-finding.

\begin{figure}[t]
    \centering
    \includegraphics[width=\linewidth]{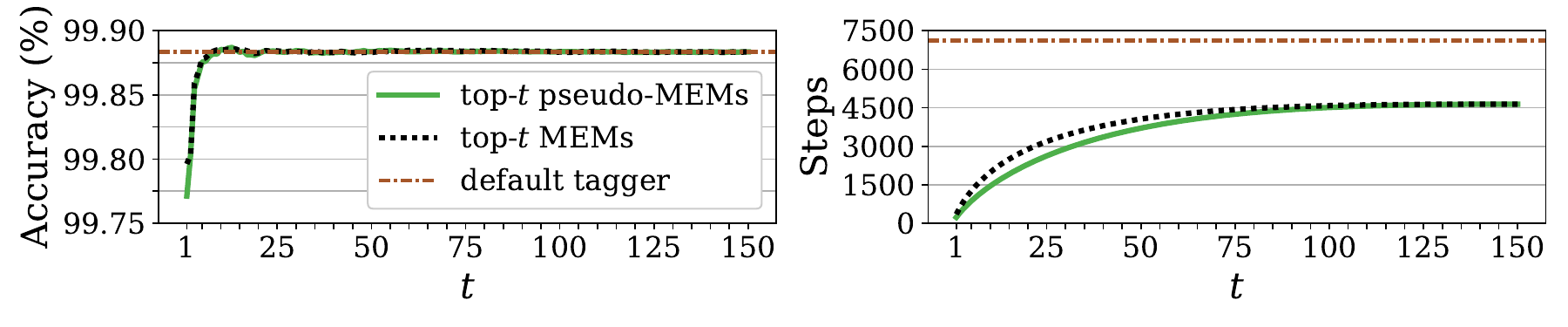}
    \caption{\tagger's accuracy {\bf (left)} and the average number of steps \bmove takes {\bf (right)} for MEM-finding.}
    \label{fig:accuracy_experiment}
\end{figure}

\begin{figure}[t]
    \centering
    \includegraphics[width=\linewidth]{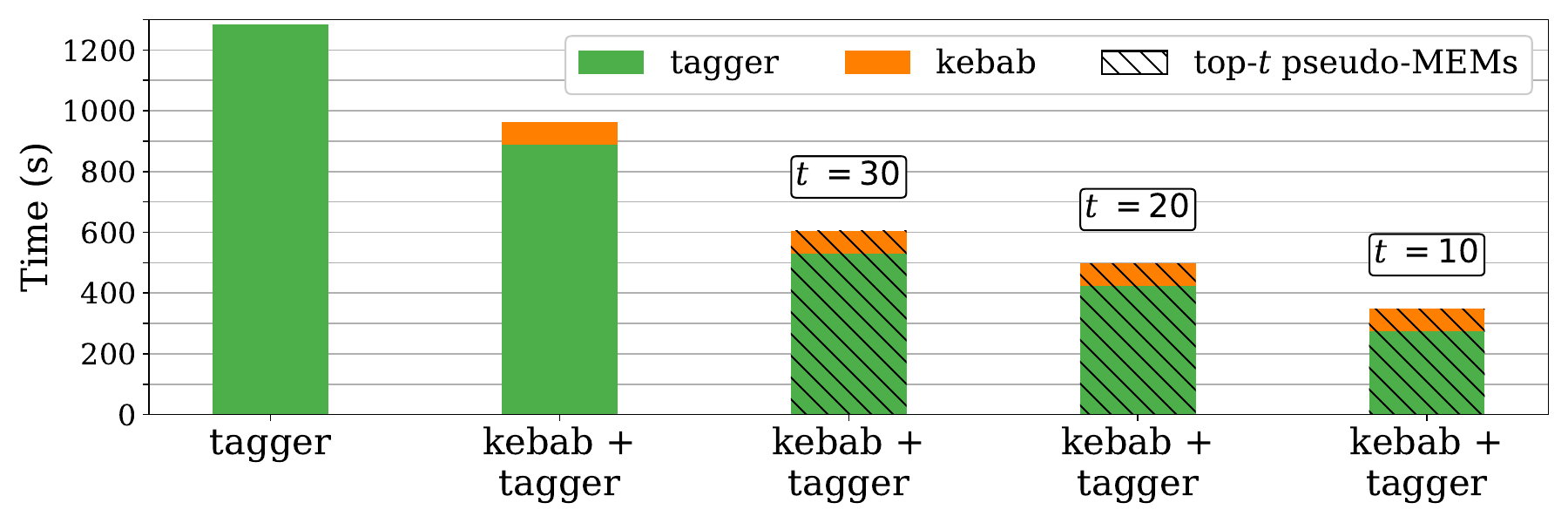}
    \caption{The time to classify using the MEMs of length at least $L = 25$ using only \tagger, or \kebab followed by \tagger, or all the MEMs of length at least $L = 25$ in the $t$ longest pseudo-MEMs in each read using \kebab followed by \tagger.}
\label{fig:multiclass_experiment}
\end{figure}

We also computed the total times, shown in Figure~\ref{fig:multiclass_experiment}, to classify the reads with \tagger after first
\begin{itemize}
\item finding all the MEMs of length at least $L$ with \bmove (``\tagger''),
\item finding all the pseudo-MEMs with \kebab and then finding all the MEMs of length least $L$ in them with \bmove (``\kebab + \tagger''),
\item finding all the pseudo-MEMs with \kebab and then finding the MEMs of length at least $L$ in the $t$ longest pseudo-MEMs from each read with \bmove (``\kebab + \tagger, $t = 30, 20, 10$'').
\end{itemize}
We ran \tagger with default settings and $L = 25$ and \kebab with $k=20$ and one hash function.  The index for \bmove took $7.869$ GB and the filter for \kebab took an additional $0.2684$ GB.  Clearly, \kebab can also speed up \tagger's pipeline.

\begin{credits}
\subsubsection{\ackname}

Many thanks to Finlay Maguire for pointing out the similarities between BML and Boyer-Moore. NKB, MZ, and BL were funded by NIH grants R01HG011392 and R56HG013865 to BL. NKB was also funded by a JHU CS PhD Fellowship and an NSERC PGS-D. LD was funded by a PhD Fellowship FR (1117322N), Research Foundation – Flanders (FWO). TG was funded by NSERC grant RGPIN-07185-2020.

\subsubsection{\discintname}

The authors declare no competing interests.

\end{credits}

\appendix

\setcounter{figure}{0}
\renewcommand\thefigure{A\arabic{figure}}

\section{Optimizations}
\label{app:optimizations}

\subsection{Bloom filter}
For a Bloom filter, let $n$ be the estimated cardinality of a set to be added, $m$ the number of bits used, $\epsilon$ the desired false positive rate, and $h$ be the number of hash functions. If the hash functions are universal, the false positive rate can be approximated to $\epsilon \approx (1 - e^{-hn/m})^{h}$. Assuming this approximation and given $n$, $\epsilon$ the minimized filter size is derived as $m=\frac{-n \cdot \ln{\epsilon}}{\ln{(2)}^2}$ with corresponding number of hashes $h=\frac{-\ln{\epsilon}}{\ln{2}}$. However, we can use fewer hashes for extra speed at the expense of a larger filter size; this is desirable for \kebab which is already small in comparison with MEM-finding indexes. Given $h$ in addition to $n$,$\epsilon$ we derive the minimized filter size as $m=\frac{-h \cdot \ln{n}}{\ln{(1-\epsilon)}^{1/h}}$.

Inserting into and querying the Bloom filter can still be slowed even using small $h$ due to the need to perform integer modulo of a hash value into the domain $[0,m)$. \textit{Fibonacci hashing} avoids an explicit modulo by ensuring that the domain size is a power of 2. Working with $64$ bits, a hash is computed by multiplying the current value by the golden ratio and then right shifting away $64 - \log_2{m}$ bits. Our implementation instead multiples by fixed seeds, but this is still universal~\cite{thorup15}. After computing our filter size $m$, we round it down to its previous power of 2, unless it is within $10\%$ of the next power of 2 in which case we round up. This can cause the false positive rate to grow but results in fast and small filters with acceptable error in our experiments.

\subsection{$k$-mer hashing}

To efficiently stream $k$-mers, we implemented the rolling nucleotide hash defined by Mohamadi et al.'s~\cite{MCVB16} ntHash. Let $rol$ be binary cyclic left rotation, $\oplus$ binary XOR, and assume single bases, e.g. $P[i]$, are replaced with a seed corresponding to the base at $P[i]$. The initial hash value is given by 
{
\small
\[
H(P[0..k-1])=rol^{k-1}(P[0]) \oplus rol^{k-2}(P[1]) \oplus ... \oplus P[k-1]
\]
}
and subsequent $k$-mers computed from the previous as 
{\small
\[
H(P[i..i+k-1])=rol(H(P[i-1..i+k-2])) \oplus rol^k(P[i-1]) \oplus P[i + k-1]
\]}
which can be seen as removing the outgoing base $P[i-1]$ and adding the incoming base $P[i + k -1]$. Let $ror$ be binary cyclic right rotation and assume $P_{c}[i]$ is the seed for the corresponding complement of the base at $P[i]$. The analogous operations for the reverse complement are 
{
\small
\[
H_{rc}(P[0..k-1])=P_{c}[0] \oplus rol(P_{c}[1]) \oplus rol^2(P_{c}[2]) \oplus ... \oplus rol^{k-1}P_{c}[k-1]
\]
}
with subsequent hashes computed as 
{
\small
\[
H_{rc}(P[i..i+k-1])=ror\!\left(H_{rc}(P[i-1..i+k-2]) \oplus P_{c}[i-1] \oplus rol^k(P_{c}[i+k-1])\right).
\]
}

Notice that, given $k$, the $rol^k$ operations required to find the next $k$-mer hash for both the forward and reverse complement can be precomputed for each base. The original ntHash paper does something similar but requires more computation to allow for flexible $k$; since our $k$ is fixed at construction time we explicitly precompute these lookup tables as well as tables for the seeds of bases and their reverse complements. This approach allows us to compute the hash value, for both the forward and reverse complement strands, of all $k$-mers by extending the previous using only lookups, XORs and a single $rol$ or $ror$ operation; we compute all hashes in linear time and fast in practice.

\subsection{Latency hiding and parallelization}

\textit{Latency hiding} avoids the time taken to load a memory word of the Bloom filter into cache by performing concurrent operations. Our approach uses it during queries by assuming a prefetch distance, set to $32$, of how many filter words we ask the CPU to fetch into memory before reading them during computation. During that time, we continue processing other hash functions/$k$-mers to find which words those require before going back to read the Bloom filter bits of queries now in cache and returning their responses.

We also \textit{parallelize} on a number of threads, giving each one read to perform concurrent operations for insertion/querying. This can change the order of when each read has its pseudo-MEMs output, but not the order of pseudo-MEMs within a read since only one thread writes at a time.

\section{Technical details}
\label{app:technical}

\subsection{Experiments}
\label{app:experiments}
Timings reported in Figures~\ref{fig:mem_length_experiment} and~\ref{fig:binary_experiment} were measured using \texttt{GNU} time on a server with an Intel(R) Xeon(R) Gold 6248R CPU running at 3.00 GHz with 48 cores and 1.5TB DDR4 memory, averaged over 10 runs using $16$ threads. Timings reported in Figure~\ref{fig:multiclass_experiment} were measured using \texttt{GNU} time on a server with an Intel(R) Xeon(R) E5-2698 v3 CPU running at 2.30 GHz with 32 cores (two threads per core) and 270 GB memory, using a single thread.

Estimating $k$-mer cardinality of the reference texts is done using $2^{20}=1048576$ bytes for HyperLogLog registers. The desired false positive rate is set to $\epsilon=1/10$. As mentioned in Section~\ref{sec:experiments}, the Bloom filter is built for $k=20$ and $h=1$ hash functions. The \kebab build command corresponding to these parameters (using 16 threads) is
$$\mathrm{\texttt{./kebab build -k 20 -e 0.1 -f 1 -t 16 [TEXT] -o [FILTER]}}$$ with the corresponding query command (using $L=40)$ $$\mathrm{\texttt{./kebab scan -o [OUTPUT] -i [FILTER] -l 40 -t 16 [PATTERN]}}$$ where \texttt{-s} is added to sort by length for top-$t$ modes. For \ropebwt and \tagger we use default flags, passing only corresponding minimum-MEM length and thread parameters.

\subsection{Output coordinates}

\begin{figure}[ht]
    \centering
    \includegraphics[width=.9\linewidth]{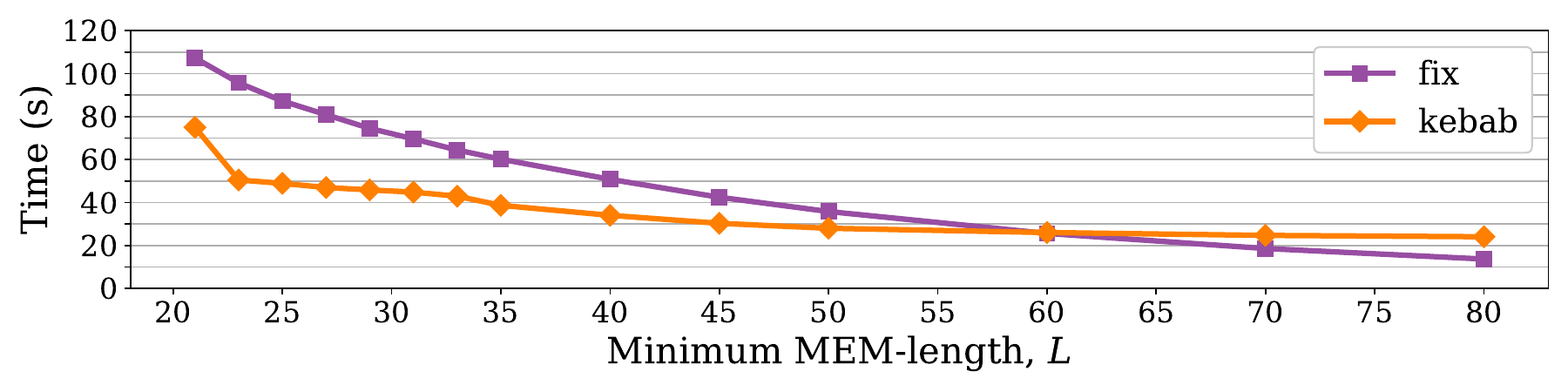}
    \caption{Compares the time to ``fix'' the output of \ropebwt using pseudo-MEMs (to that of \ropebwt alone) against the \kebab filter step. Where \kebab's speed depends only on the pattern lengths, fixing output depends on the number of distinct MEMs.}
    \label{fig:fix_time}
\end{figure}

Passing pseudo-MEMs to \ropebwt results in a slight variation in the output coordinates, since it reports the positions of MEMs with respect to the given input pattern which are no longer full reads; however, pseudo-MEMs are output with \texttt{[SEQ]:[START]-[END]} identifiers to relate them back to their original pattern. Thus, a script can optionally be run to ``fix'' this output to exactly match that of running \ropebwt alone by reorienting pseudo-MEM coordinates back to full pattern coordinates. This does not change the actual MEMs found and they are still recoverable from the pseudo-MEM files, so we omit this step in Figures~\ref{fig:mem_length_experiment} and~\ref{fig:binary_experiment}. The run-time of just \kebab (with data/parameters of Figure~\ref{fig:mem_length_experiment} and corresponding settings from Section~\ref{app:experiments}) compared to running our simple, single-threaded fix is shown in Figure~\ref{fig:fix_time}.

\end{document}